\documentclass[twocolumn,showpacs,prl,superscriptaddress,footnoteinbib]{revtex4}
\usepackage[pdftex]{graphicx}
\usepackage{amsfonts}
\usepackage{amsmath}
\usepackage{amssymb}
\usepackage{color}
\usepackage{hyperref}

\newcommand{\vs}{v_{\textrm{STM}}}

\newcommand\threed{{\rm{3D}~}}

\newcommand\tid{{\textrm{TI}}}
\newcommand\ti{{\textrm{TI}~}}

\newcommand\arpesd{{\textrm{ARPES}}}
\newcommand\arpes{{\textrm{ARPES}~}}

\newcommand\tmrd{{\rm{TMR}}}
\newcommand\tmr{{\rm{TMR}~}}

\newcommand\stmd{{\rm{STM}}}
\newcommand\stm{{\rm{STM}~}}

\newcommand\spstm{{\rm{SPSTM}~}}

\newcommand\trsd{{\rm{TRS}}}
\newcommand\trs{{\rm{TRS}~}}

\newcommand\dosd{{\rm{DOS}}}
\newcommand\dos{{\rm{DOS}~}}

\begin{document}

\title{Tunnel  Magentoresistance  scan of  a pristine three-dimensional topological insulator}
\author{Sthitadhi Roy}
\author{Abhiram Soori}
\affiliation{Max-Planck-Institut f\"ur Physik komplexer Systeme, N\"othnitzer 
Stra$\beta$e 38, 01187 Dresden, Germany}
\author{Sourin Das}
\affiliation{Department of Physics and Astrophysics,
University of Delhi, Delhi 110 007, India}
\affiliation{Max-Planck-Institut f\"ur Physik komplexer Systeme, N\"othnitzer 
Stra$\beta$e 38, 01187 Dresden, Germany}

\begin{abstract}
Though the Fermi-surface of surface states of a \threed topological insulator (\tid) has zero magnetisation, an arbitrary segment of the full Fermi surface has a unique magnetic moment consistent with the type of spin-momentum locking in hand. We propose a three-terminal set up which directly couples to the magnetisation of a chosen segment of a Fermi surface hence leading to a finite tunnel magnetoresistance (\tmrd) response of the non-magnetic \ti surface states,  when coupled to spin polarised \stm probe. This multi-terminal \tmr not only provides an unique signature of spin-momentum locking for a pristine TI but also provides a direct measure of momentum resolved out of plane polarisation of hexagonally warped Fermi surfaces relevant for $Bi_2Te_3$ which could be as comprehensive as spin resolved \arpesd. Implication of this unconventional \tmr is also discussed in the broader context of 2-D spin-orbit (SO) materials.
\end{abstract}
\maketitle
\begin{figure}
\begin{center}
\includegraphics[width=.8\columnwidth]{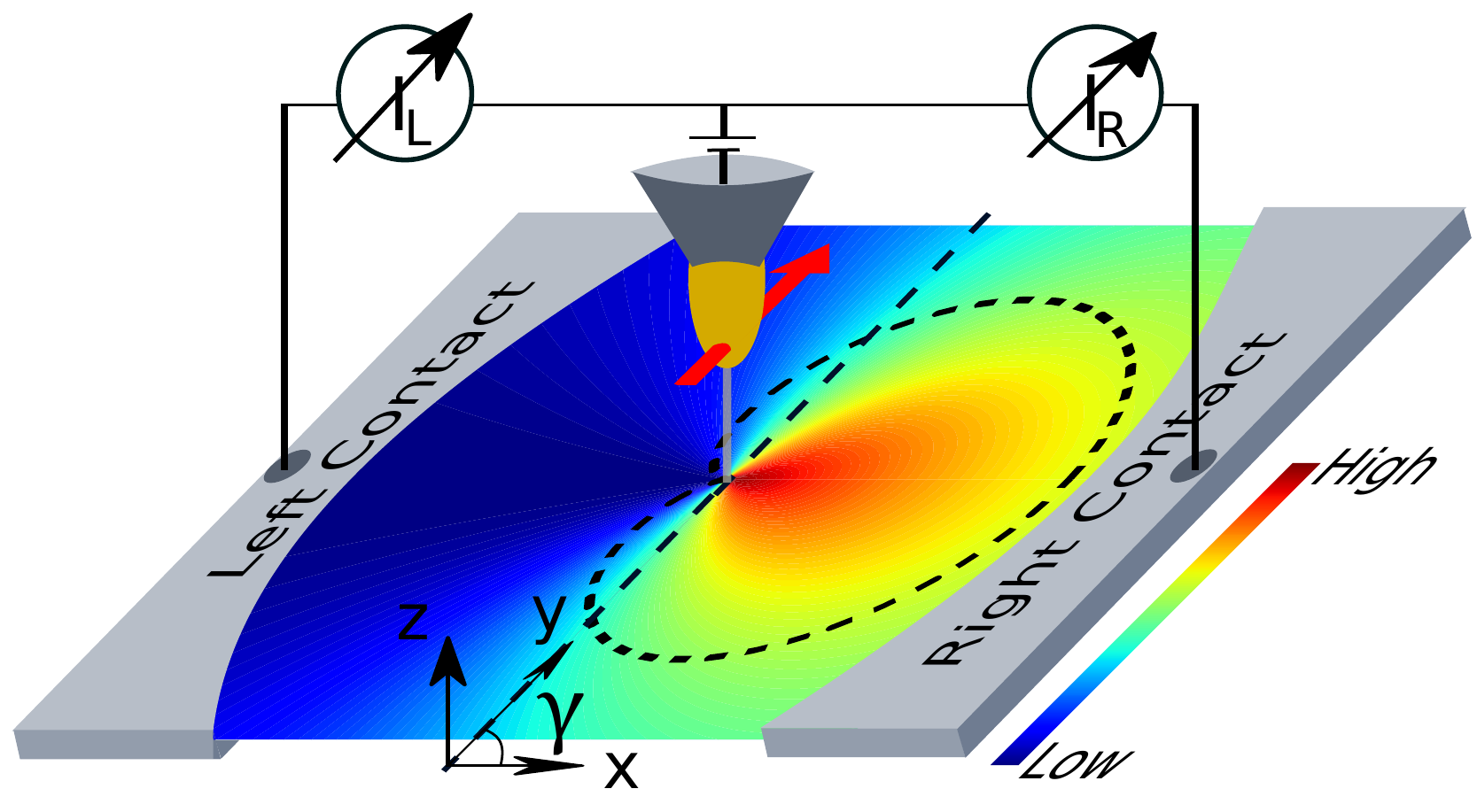}
\end{center}
\caption{A schematic of the setup. The colour density shows the current density profile on the surface of the TI consistent with the polarisation direction of the tip shown by red arrow pointing along $y$-axis and the  spin-momentum locking angle, $\theta_{\textrm{L}}=\pi/2$. Dotted curve is a polar plot of the tunnelling current amplitude at the point of injection showing the profile of the current anisotropy. $I_{L/R}$ are the current (in arbitrary units) carried by the left and right contact.}
\label{fig:schematic}
\end{figure}
\begin{figure}
\begin{center}
\includegraphics[width=0.8\columnwidth]{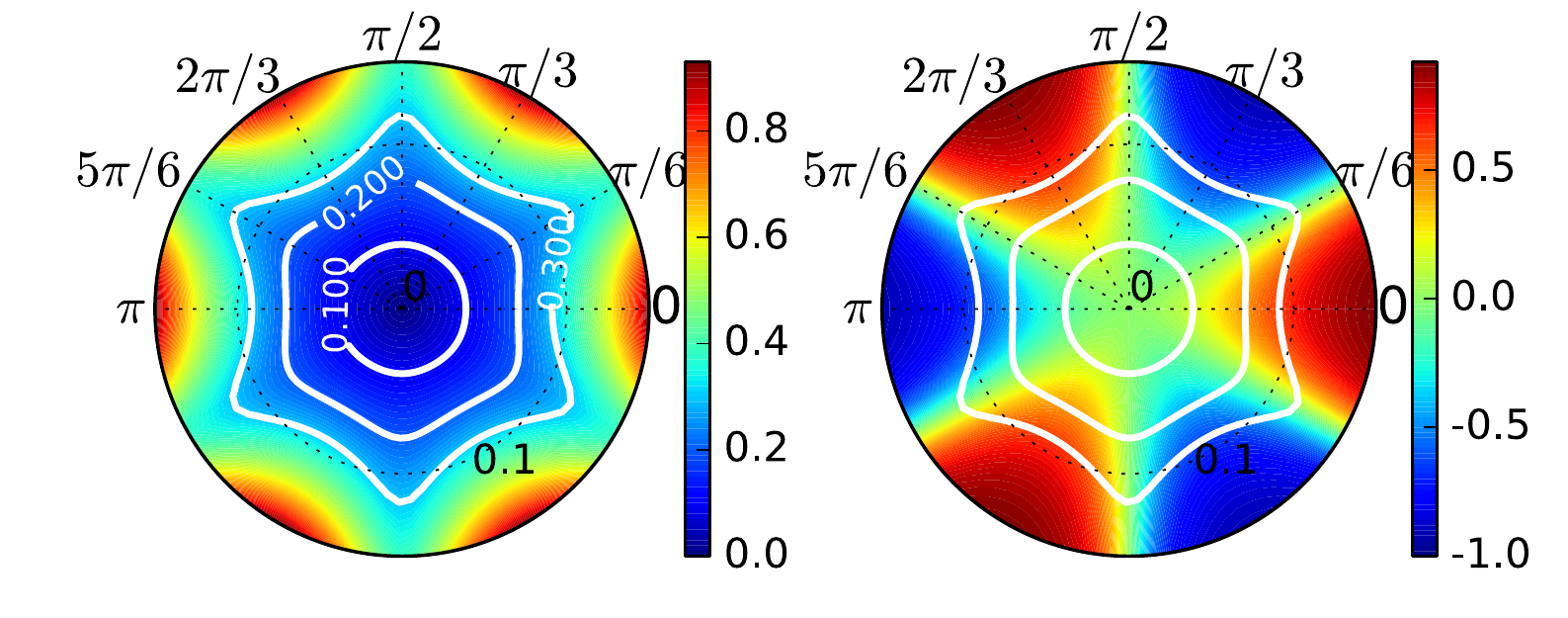}
\end{center}
\caption{The energy spectrum in units of eV (left) and the out of plane magnetisation (right) in the momentum space for the Hamiltonian with the warping term. The white contours depict the Fermi surface at $E_F=0.1, 0.2 \text{ and } 0.3$ eV respectively.The momentum magnitudes in the polar plots are in units of $\text{\AA}^{-1}$.}
\label{fig:warp_spect_sz}
\end{figure}
\begin{figure}[t]
\begin{center}
\includegraphics[width=0.8\columnwidth]{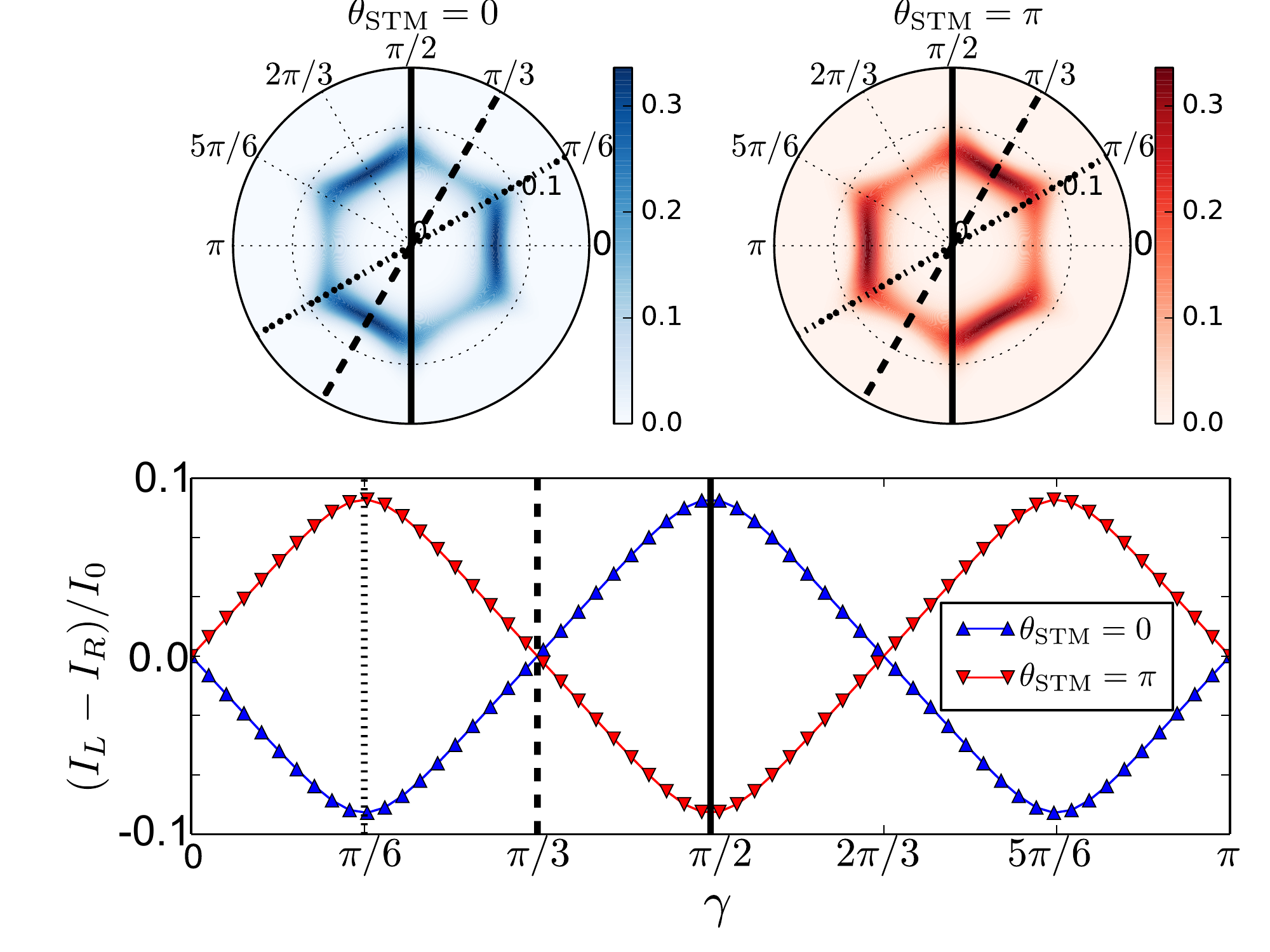}
\end{center}
\caption{Top: The current distributions in the momentum space with the \stm magnetisation being up (left) and down (right). The momentum magnitudes in the polar plots are in units of $\text{\AA}^{-1}$. Bottom: $\Delta I$ as a function of $\gamma$. The three different vertical lines at $\pi/6$, $\pi/3$ and $\pi/2$ correspond to the three different bipartition of the top panel (corresponding line styles).}
\label{fig:cur_asym_warp} 
\end{figure}

 {{\sl {\underline{Introduction}:}}} 
The two popular  probes used to scan the surface states of \cite{kane,zhang} 3-D TIs are spin polarised \arpesd\cite{Hasan} or \stmd\cite{Yazdani, Tong}. Spin polarised \arpes seems to have an edge over the \stm  as it couples more directly to the spin texture of the Fermi surface. In this paper we propose a multi-terminal set up involving spin-polarised \stm (SPSTM) which directly couples to the spin texture of the Fermi surface leading its straightforward read readout . This read out is theoretically understood in terms of a new kind of \tmr\cite{PhysRevB.39.6995} response between the magnetised \stm and the non-magnetic \ti surface which relies crucially on its multi-terminal character as described below. \\
{{\sl{\underline{Proposed set up}:}}} 
The  proposed set up comprises of  two contact pads placed diametrically opposite to each other on the surface of the \ti while electrons are injected from the \spstm placed at the centre of the sample as shown in Fig.~\ref{fig:schematic}. The surface can be imagined to be divided into two halves by a line through the centre of the sample perpendicular to the direction joining the two contacts, (for future reference, we mention that the angle made by this partitioning line with the x-axis is denoted by $\gamma$, see Fig.~\ref{fig:schematic}). Each contact measures the current flowing in the surface in its own half. We show that the total current $I_{0}=I_L + I_R$ is insensitive to current anisotropy discussed above owing to zero magnetisation of Fermi surface but $\Delta I=I_L - I_R$ is very sensitive to the current anisotropy and leads to a finite \tmr response with the spin polarised \stm which oscillates as a function of $\gamma$. Note that $\gamma$ can be changed simply by rotating the sample with respect to 
the tip about $z$-axis. 
We show that $\Delta I$ measured as a function of  $\gamma$  leads to a direct reconstruction of in-plane spin texture in the momentum space, i.e., we can extract the angle of spin-momentum locking ($\theta_{\textrm{L}}$) and the chirality; left chiral or right chiral from this study. Then we extend the calculation to include warping effects and show how a 
similar measurement can uniquely reconstruct the six fold symmetric out-of-plane spin texture of the warped Fermi surface too.\\
{{\sl {\underline{Model}:}}} 
 We start with the generic Hamiltonian for the \threed \ti surface state given by 
\begin{eqnarray}
\mathcal{H}_{\textrm{TI}} &=& \hbar v_{\textrm{F}} \sum_{\vec{k}}\Psi_{\textrm{TI}}^\dagger({\vec{k}}) (\vec \sigma \times \vec k)_z \Psi^{}_{\textrm{TI}}({\vec{k}})\,,
\label{eq:hamiltonian_TI} 
\end{eqnarray}
where $\Psi_{\textrm{TI}}^{} ({\vec{k}}) = {{1}/{\sqrt{2}}} (1,e^{i(\phi_{\vec{k}} + \theta_{\textrm{L}})})^T\hat{c}_{\vec{k}}$ and $\hat{c}_{\vec{k}}$ is the annihilation operator with momentum $\vec{k}$. $\phi_{\vec{k}}=\tan^{-1}(k_y/k_x)$ is the polar angle of the momentum vector. For Eq.~(\ref{eq:hamiltonian_TI}), the spin-momentum locking angle $\theta_{\textrm{L}}$ is $\pi/2$, consistent with popularly known 
\ti materials\cite{Zhang1}. Assuming a flat density of states, a model Hamiltonian for a fully spin-polarised \stm tip is written as $\mathcal{H}_{\textrm{STM}}=\sum_{\vec{k}} \varepsilon_{\vec{k}}  \hat{d}_{\vec{k}  ,\uparrow}^\dagger \hat{d}^{}_{\vec{k}, \uparrow}$, where the \stm electron annihilation operator in real space is given by $\Psi_{\textrm{STM}}({\vec{r}}) = \int d \vec{k} e^{i \vec{k}\cdot\vec{r}} ( \cos(\theta_{\textrm{STM}}/2), \sin(\theta_{\textrm{STM}}/2) e_{}^{i\phi_{\textrm{STM}}})^T \hat{d}^{}_{\vec{k},\uparrow}$ and $\varepsilon_{\vec {k}}=v_\textrm{STM} \vert \vec{k} \vert$ where $v_\textrm{STM}$ is the Fermi velocity of the \textrm{STM} and $\theta_{\text{STM}}(\phi_{\text{STM}})$ is the polar (azimuthal) angle of the STM spin.
The \stm tip is assumed to be weakly coupled to the \ti surface by a tunnelling Hamiltonian, 
 $\mathcal{H}_{\textrm{tunn}} = J (\Psi_{\textrm{TI}}^\dagger(\vec{r}=0)\Psi^{}_{\textrm{STM}}(\vec{r}=0) + \textrm{h.c.})$ which upon Fourier transforming gives  $\mathcal{H}_{\textrm{tunn}} =J \sum_{\vec{k},\vec{k'}} z_{\vec{k}} ^{} \hat{c}_{\vec{k}}^\dagger \hat{d}^{}_{\vec{k'},\uparrow} + \textrm{h.c.}$, where
\begin{eqnarray}
z_{\vec{k}}^{} &=& \frac{1}{\sqrt{2}}\left(\cos\frac{\theta_{\textrm{STM}}}{2}-i\sin\frac{\theta_{\textrm{STM}}}{2}e^{i(\phi_{\textrm{STM}} -\phi_{\vec{k}})}\right)\,,
\label{eq:zk}
\end{eqnarray}
has the information about the overlap of the \stm spinor and the \ti spinor for each momentum mode $\vec{k}$ which ultimately decides how much injection happens in each $\vec{k}$ and hence the current anisotropy discussed above.
The current operator is defined as $\hat{I} = d\hat{N}_{\textrm{STM}}/dt = \frac{i}{\hbar}[\mathcal{H}, \hat{N}_{\textrm{STM}}]$
where $\hat{N}_{\textrm{STM}} = \int d \vec{k} \hat{d}_{\vec{k},\uparrow}^\dagger \hat{d}_{\vec{k} ,\uparrow}^{}$ and $\mathcal{H} = \mathcal{H}_{\textrm{TI}}+\mathcal{H}_{\textrm{STM}}+\mathcal{H}_{\textrm{tunn}}$. The expectation value of the current at some time $t$ is given by
$\langle \, I (t) \, \rangle= \langle \, G  \, \vert \, e^{i\mathcal{H}t } \, \hat{I} \, e^{-i\mathcal{H}t} \, \vert \, G \, \rangle$,
where $\vert \, G \, \rangle$ is the ground state  of $\mathcal{H}$ in presence of a chemical potential bias ($\mu_{\text{STM}}\ne\mu_{\text{TI}}$) between STM and TI. Treating  $\mathcal{H}_{\textrm{tunn}}$  perturbatively, we get  
\begin{eqnarray}
\langle \, I (0) \, \rangle &=& \dfrac{i}{\hbar}\int_{-\infty}^{0}dt' \langle \, g \, \vert [\mathcal{H}_{\textrm{tunn},I}(t'),\hat{I}_{I}(0) ] \vert\, g\, \rangle \, ,
\label{eq:curr2}
\end{eqnarray}
where the $I$ in the subscript stands for the interaction picture and  $\vert \, g \, \rangle = \vert \,g \, \rangle_{\textrm{TI}}\otimes \vert g \rangle_{\textrm{STM}}$ is the non-interacting ground state with $\mathcal{H_{\textrm{tunn}}}$ is treated as interaction. Decomposing the current in momentum space we obtain,  $\langle \, I  \, \rangle = \frac{e}{\hbar}\int d\vec{k} \int d {\vec k}^\prime \vert z_{\vec{k}}\vert^2\chi_{\vec{k},{\vec k}^\prime}\delta(\varepsilon_{\textrm{STM}}({\vec k}^\prime)-\varepsilon_{\textrm{TI}}(\vec{k}))$, where $\vert z_{\vec{k}}\vert^2$ (Eq. (\ref{eq:zk})) has the information of the spinor overlaps,  $\chi_{\vec{k},{\vec k}^\prime} = \int_{-\infty}^{\infty} d\tau~ \textrm{Im}[{G}_{\textrm{TI}}(\vec{k},0;\vec{k},\tau){G}_{\textrm{STM}}({\vec k}^\prime,\tau;{\vec k}^\prime,0)]$ has the Green's functions, where ${G}$ denotes the standard time ordered Fermionic Green's functions and the delta function ensures energy conservation. Hence we obtain a momentum resolved current 
given by 
\begin{eqnarray}
\langle \, I (\vec{k}) \, \rangle &=& \dfrac{e}{\hbar} \vert  z_{\vec{k}}  \vert^2 \chi_{\vec{k}}\,,
\label{eq:curr_k}
\end{eqnarray} 
where $\chi_{\vec{k}} = \int_{-\infty}^{+\infty} d {\vec k}^\prime  \chi_{\vec{k},{\vec k}^\prime} \delta(\varepsilon_{\textrm{STM}}({\vec k}^\prime)-\varepsilon_{\textrm{TI}}(\vec{k}))$. 
The total tunnelling current is just a sum over $\langle \, I(\vec{k}) \, \rangle$ for all possible $\vec{k}$ living in the bias window.
The angular distribution of the injected current in real space is same as the angular distribution of momentum resolved current about the tunnelling point. As a consistency check for this, we evaluate the expectation value of the current vector operator for a pristine \ti surface $\hat{\vec{j}}(\vec{r}) = (\sigma^y(\vec{r}),-\sigma^x(\vec{r}))$ at $\vec{r}=0$  perturbatively to second order in tunnel Hamiltonian and obtain $\langle \, \hat{\vec{j}}(\vec r=0) \, \rangle = \sum_{\vec{k}} \langle \, 
I(\vec{k}) \, \rangle \hat{n}_{\vec k}$ which reconfirms our interpretations of  $\langle \, I(\vec{k})\, \rangle $ being the real space angular distribution of current about the injection point. Here $\hat{n}_{\vec k}$ is a unit vector pointing along $\vec k$. Owing to the azimuthally symmetric (in $k-$space) Fermi surface, $\langle \, I(\vec{k}) \, \rangle$, turns out to be separable in its dependence on $\vert \vec{k} \vert$ and $\phi_{\vec{k}}$ as $\langle \,I (\vec{k}) \, \rangle=I_{\vert\vec{k}\vert}I_{\phi_{\vec{k}}}$ with
\begin{eqnarray}
I_{\phi_{\vec{k}}}  &=& 1 + \sin\theta_{{STM}}\sin(\phi_{{STM}} - \phi_{\vec{k}})\,.
\label{eq:cur_profile}
\end{eqnarray}
It is clear from the above result that the total injected current which is obtained by summing over all possible momenta in the bias window leads to a  current which is independent of the direction of \stm tip magnetisation\footnote{In Ref.\cite{saha} using Bardeen tunnelling formula a finite \tmr response was found for two terminal case which was then set to zero based on symmetry arguments, but in present approach this is natural out come of the formulation.}  (zero \tmrd) consistent with \trs in \tid. But the current asymmetry defined as  $\Delta I=I_L-I_R=(\int_\gamma^{\gamma+\pi}-\int_{\gamma+\pi}^{\gamma+2\pi})d\phi_{\vec{k}} I_{\phi_{\vec{k}}} \int \vert \vec k \vert d\vert\vec{k}\vert I_{\vert\vec{k}\vert}$ which could be measured directly in the set up depicted in Fig.~\ref{fig:schematic} shows a finite \tmr as function of $\gamma$ given by
\begin{eqnarray}
\Delta I &=& \dfrac{4e J^2}{ \hbar^4 v_f^2 \vs} \mathcal{F} \cos(\phi_{{STM}} - \gamma) \sin\theta_{{STM}}\,,
\label{eq:asym_final}
\end{eqnarray}
where $\mathcal{F}=\int_{-\infty}^{+\infty}dE~E(\text{n}_\text{F}^{\text{TI}}-\text{n}_\text{F}^{\text{STM}})$ is obtained from the $\vert\vec{k}\vert$ integral by appropriately putting in the density of states, where $\text{n}_\text{F}(E,\mu,T)$ denotes the Fermi function.
\begin{figure}
\begin{center}
\includegraphics[width=0.8\columnwidth]{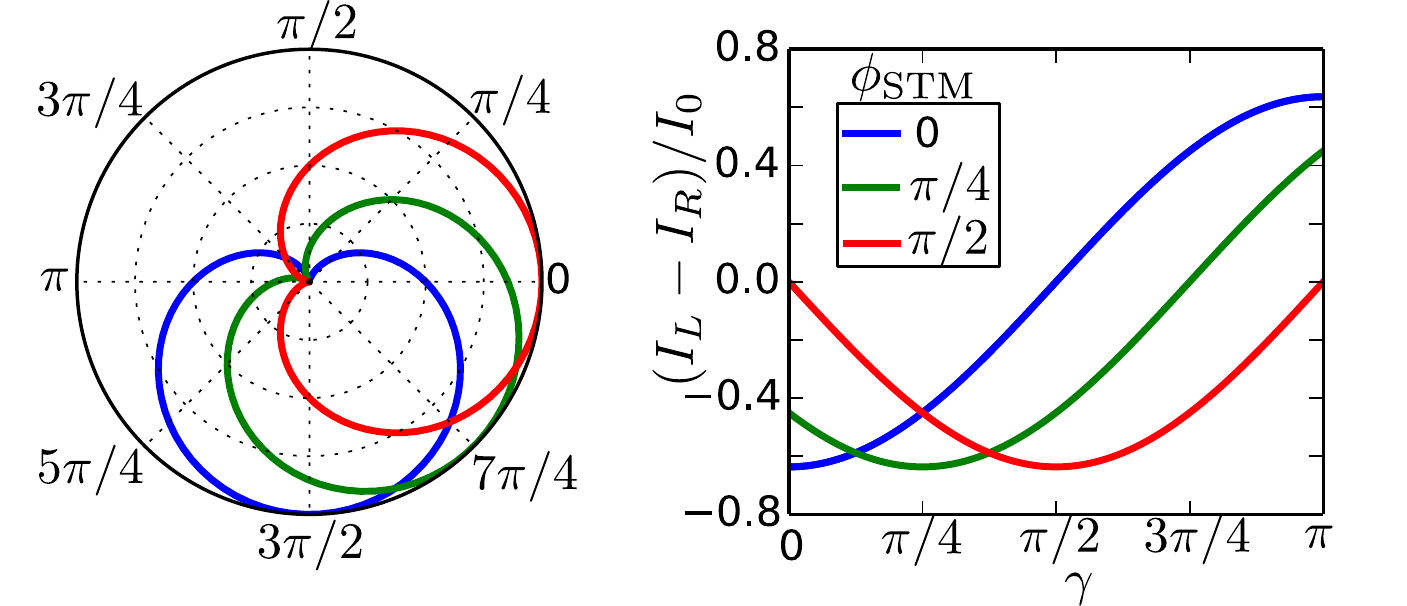}
\end{center}
\caption{Left:The angular profile of the current as obtained from Eq.(\ref{eq:cur_profile}) for $\theta_{{STM}}=\pi/2$ and different values of $\phi_{{STM}}$ as mentioned in the legend. Right: The resulting current asymmetries as function of $\gamma$ as obtained from Eq.~(\ref{eq:asym_final}).}
\label{fig:prof_nowarp}
\end{figure}
 {{\it {\underline{Reconstructing spin texture}:}}} 
We will now demonstrate that the above expression can be directly exploited to uniquely identify the spin-momentum locking angle and the chirality. At this point it is important to note that all the in-plane angles like $\gamma$ and $\phi_{{STM}}$ are measured with respect to the positive direction of $x$-axis along the anti-clock wise direction. For the case of ${\phi_{{STM}} =\pi/2}$  observing a zero in $\Delta I$ at $\gamma=0$ (see Fig.(\ref{fig:prof_nowarp})) implies that the momentum modes pointing towards the left and right contact starting from the tip position have a locked-spin which is pointing perpendicular to the STM polarisation pointing along $y$-axis (see  Fig.~\ref{fig:schematic}) so that the injected current gets symmetrically distributed between left and right contact. Assuming a planar spin texture,  it directly tells us that the spin momentum locking angle $\vert \theta_{\textrm{L}} \vert =\pi/2$. Of course this conclusion relies on the assumption that the Fermi-surface is circular in 
shape so that spin on each half can be added up symmetrically . Now we are left with two possibilities; the two oppositely directed momentum modes discussed above pointing towards left and right contact have a locked-spin either pointing parallel and anti-parallel to the $x-$axis or the other way round respectively. And this information is nothing but the spin chirality of the Fermi surface. To settle the chirality, we observe that $\Delta I$ is maximally negative for $\gamma=\pi/2$. This implies that maximal share of the injected current is flowing to the right contact (as depicted in Fig.~\ref{fig:schematic}) implying that the momentum mode pointing towards right starting from the tip position has a locked-spin which is parallel to the tip magnetisation direction. Hence the study of $\Delta I $  also implies that the momentum mode pointing along x-axis has a locked-spin pointing along y-direction hence reading out the spin chirality of the Fermi surface in hand.  So its leads to conclusion that the spin-momentum locked spin is uniquely given by  $<\vec{\sigma}(\vec{k})>=(-\sin\phi_{\vec{k}},\cos\phi_{\vec{k}})$. Hence our claim of reconstructing the Fermi-surface spin texture using the proposed three-terminal \tmr data is clearly demonstrated.  
 For an arbitrary spin-momentum locking angle $\theta_{\textrm{L}}$, $\Delta I\sim\sin(\gamma-\phi_{\textrm{STM}}+\theta_{\textrm{L}})$; the maxima in its magnitude occurs at $\gamma = \phi_{\textrm{STM}}-\theta_{\textrm{L}}+\pi/2$ and hence the spin-momentum locking angle can be extracted. The sign of the first maxima of  $\Delta I$ as we increase $\gamma$ from zero gives the chirality.\\   
 {{\sl { \underline{TMR and $I_{L/R}$}:}}}  The next step is to recast $I_{L/R}$ or equivalently $\Delta I$  in an explicit \tmr form\cite{PhysRevB.39.6995} which puts our idea on a firm footing and adds further transparency to the discussion above. Note that a net spin polarisation vector can be obtained by performing a vector sum of spin polarisations of each momentum mode living on half of the Fermi surface of \ti surface states, where the Fermi surface bipartition is done along the line defining $\gamma$.  This quantity for each half of the Fermi Surface is given by $\vec{S}_{{half,L/R}}(\gamma,\mu) =\mp (\rho^{TI}_{\mu}/\pi)(\cos\gamma \hat{x}+\sin\gamma\hat{y}) $ which sums to zero due to \trsd. Here $\rho^{TI}_{\mu}=2 \pi \mu/(\hbar v_F)^2$ is the density of states (\dosd) of \ti at chemical potential $\mu$ where $\mu$ is measured from the neutrality point. Then, by extending  Eq.~(\ref{eq:asym_final}) to include finite polarisation of the \stm tip, in linear response limit we obtain, 
\begin{eqnarray}
I_{L/R} = 
\frac{  \pi J^2 e^2  \rho_{\mu}^{{TI}} \rho^{{STM}}}{ h} [1\mp \frac{2 p}{\pi}  \hat {S}_{{half}}(\gamma,\mu)\cdot \hat {S}_{{STM}}] \,V
\label{eq:asym_final1}
\end{eqnarray}
where $\rho^{STM}$ is the spin averaged \dos of the tip, ${p}$ is the polarisation of the tip given by $(\rho_{\uparrow}^{STM}-\rho_{\downarrow}^{STM})/(\rho_{\uparrow}^{STM}+\rho_{\downarrow}^{STM})$ 
and $V$ is the applied voltage bias between tip and TI. $\hat S_{{half}}$ and  $\hat {S}_{{STM}}$ are unit vectors along $\vec{S}_{{half}}$ and magnetisation direction of the tip. We see that indeed the left and right contact shows a standard  \tmr response\cite{PhysRevB.39.6995}  having opposite signs (due to \trsd) with the magnetised \stm and the pure magnetic response can be extracted from it simply by taking an anti-symmetric combination of the two which is nothing but $\Delta I$. Hence spin-momentum locking together with multi-terminal set up leads to this exotic situation where large \tmr response is extracted out of a non-magnetic material which is shown to be useful for characterising the material itself. This is the central finding of this paper. It is important to the note that the expression for $I_{L/R}$ obtained above remains valid as long as spin-momentum locking is protected by \trs independent of the fact that the spin lives in the x-y plane or has components along the z-direction. The only 
necessary change in  Eq.~(\ref{eq:asym_final1}) is,  we need to replace $\vec{S}_{{half,L/R}}(\gamma,\mu)$ for the present case with its corresponding counterpart.  This observation is crucial for application of Eq.~(\ref{eq:asym_final1})  to the next case (warped case) where we have out of plane spin-momentum locking.\\
{{\sl { \underline{Warping case}:}}} 
Next, we consider a hexagonal warping term\cite{PhysRevLett.103.266801} in the Hamiltonian of the \ti  given by $
\mathcal{H}_{\textrm{W}}  = \sum_{\vec{k}}\frac{\lambda}{2}(k_+^3 + k_-^3) \sigma^z;~\mathcal{H}_{\text{TI}}=\hbar v_F\sum_{\vec{k}} (\vec{\sigma} \times \vec{k})_z + \mathcal{H}_{\textrm{W}}$, which gives rise to a spin texture on the Fermi surface whose out-of-plane polarisation has a staggered structure in the $\vec{k}$- space with six symmetric regions of alternating positive and negative out-of-plane magnetisation (see Fig.(\ref{fig:warp_spect_sz})).
The increasing strength of the TMR signal (with STM magnetization pointing along the $z$- direction) with increased doping of the TI surface state provides a direct  measure of the magnitude of out-of-plane magnetization of the Fermi surface. The TMR signal strength is observed to grow monotonically with the chemical potential of TI surface state measured from the neutrality point.\\
{{\sl {Reconstructing out of plane spin texture:}}}  In this case we keep the \spstm magnetisation either parallel or anti-parallel to the z-axis.  Electrons injected from spin polarised \stm pointing along positive z-direction see six alternating regions with positive and negative out-of-plane polarisation  on the Fermi surface leading to alternating high and low overlap of the wavefunctions in momentum space. This leads to a high current density in three regions and low current density in the rest of the three. This pattern reverses itself as the \stm magnetisation is reversed. This pattern of the current distribution is evaluated using Eq.~(\ref{eq:asym_final1}) and is shown in Fig.~\ref{fig:cur_asym_warp}. This current pattern can be measured by the same three-terminal current measurements setup defined above but there is a crucial difference. 
In the previous case rotating the orientation of the contacts (i.e. changing $\gamma$)  was equivalent to rotating the sample about z-axis owing to azimuthal symmetry of the  Fermi surface but in this case the contacts are to be engineered separately for each $\gamma$ while keeping the orientation of underlying lattice on the x-y plane fixed as the hexagonal warping of the spectrum is a direct reflection of the underlying lattice. Hence a minimal set up requires multiple samples with identical orientation of underlying lattice in the x-y plane with contact pads rotated with respect to each other.  The 3-fold symmetry of the lattice, which gives rise to the 3-fold symmetry of the out of plane spin polarisation causes a periodicity of $2\pi/3$ in $\Delta I$ measured as a function of $\gamma$  (see Fig.~\ref{fig:cur_asym_warp}). It can be seen from the figure that if the line partitioning the sample into two halves goes through the corners of the hexagonal pattern of the Fermi surface, then the asymmetry is 
extremal, on the other hand if it goes through the centre of the sides of the hexagon, the asymmetry is zero.  A periodicity of $2\pi/3$ in $\Delta I$  as we vary $\gamma$ implies that there is a 3-fold symmetry in the texture. Three successive peaks (dips) turning into dips (peaks) as we flip the \stm magnetisation direction from positive $z$-axis to negative z-axis implies the presence of three regions with positive out-of-plane magnetisation and three others with negative. The symmetry in magnitude of the dip and peak indicates the up and down polarised  patches being equal in length on the Fermi surface. This demonstrates that $\Delta I$ could be used for uniquely reconstructing  the out of plane polarisation of the Fermi surface in the warped case. \\
{{\sl {\underline{Experimental feasibility}:}}} Implementation of our proposed set up requires the \ti surface state to be in ballistic limit. $Bi_2Te_3$, a 3D TI material has a reported elastic mean free path of $235 nm$\cite{Qu2010} and hence a micron by micron sized sample could be considered reasonable and is an experimentally feasible sample size\cite{Yazdani}. As our proposal strongly depends on the spin degree of freedom, spin-relaxation length is also a very relevant length scale. Though there are no clear estimates for this quantity, it is clearly bounded from below by the elastic mean free path\cite{Pesin2012} in the absence of magnetic impurities. Our proposal relies strongly on the existence of almost reflection-less contacts. This fact could be of concern because such reflections could suppress the amplitude of the proposed current asymmetry, however the important point to note here is the fact that a contact should be largely reflection-less due to the spin-momentum locked nature of the surface 
states. Any reflection from the contact will require flipping the spin and hence will be suppressed for non-magnetic contacts. This indeed seems to be the case from experimental results reported in Ref.\cite{Li} where contacts with resistance as low as $1\, m \Omega$ were implemented.\\
{{\it {\underline{Discussion and outlook}:}}} In this paper we have shown that a  \tmr response can be obtained even from a non-magnetic material like a \ti surface when weakly coupled to a ferromagnet, provided we subscribe to a multi-terminal set up. Proximity induced magnetism in a non-magnetic material has been used to study TMR in earlier works\cite{PhysRevB.81.121401,PhysRevB.89.085407}, however in our proposal the weak coupling ensures that the response obtained is indeed of the pristine non-magnetic material which gives it a clear advantage over its predecessors. The fundamental physics in our proposal which is responsible for extracting \tmr response from a non-magnetic material is spin-momentum locking which is generic to any strong SO coupled material, for example a 2D Rashba SO coupled material which has two circular Fermi surfaces with their individual spin-momentum locking akin to that of a strong 3D TI surface, except for, the relative chirality of the spin texture between the two Fermi surfaces of the Rashba material being opposite to each other. Hence, a \tmr response can be extracted as before by calculating $\Delta I/I_0$ in a set up identical to Fig.(\ref{fig:schematic}) using our formulation which turns out to be
\begin{equation}
\frac{\Delta I}{I_0} =  \delta\rho~ {\bigg{\{}} (\frac{2p}{\,\pi}\,) ~\hat S_{half}(\gamma,E_F)\cdot\hat{S}_{STM} \bigg{\}},
\end{equation}
where $\delta\rho$ is given by $({\rho_{\text{out}}-\rho_{\text{in}}})/({\rho_{\text{out}}+\rho_{\text{in}}})$, $\hat{S}_{half}$ is the unit vector pointing along the total magnetisation of half of the Fermi surface of the outer Fermi surface for Rashba spectrum and $\rho_{\text{out(in)}}$ is the DOS for outer (inner) Fermi surface for the Rashba Hamiltonian. Note that the TMR response for the 2D Rashba material is similar to that of the TI apart from the suppression factor of $\delta\rho$ which could be optimised by varying $E_F$.  A similar calculation for the case of arbitrary 2D spin orbit Hamiltonian is straight forward though could be tedious if the Fermi surfaces are not circular.  Hence this new concept of multi-terminal  \tmr introduced in this article is of direct relevance for studying Fermi surface topology and spin texture of arbitrary 2D spin orbit materials in general.  \\
{{\it {Acknowledgements:}}} It is a pleasure to thank Y. Gefen, A. Junck,  Y. Oreg,  S. Rao and K. Sengupta for discussions, and J. H. Bardarson, K. Saha and D. Sen for discussions and critical reading of the manuscript, S. Patnaik and P. Sen for discussion on experimental feasibility and  M. Deshmukh for pointing out Ref.\cite{Li}. 

\bibliographystyle{apsrev}

\bibliography{references}
\end{document}